\documentclass[aps,prl,twocolumn,showpacs,letterpaper,showpacs,superscriptaddress]{revtex4-1}
\usepackage{graphicx,amsmath,amssymb,amsfonts,latexsym,xcolor,dcolumn,bm,epsfig,subfigure}
\usepackage[plainpages=false,hyperfootnotes=false,colorlinks=false]{hyperref}
\usepackage[normalem]{ulem}
\usepackage{dsfont}

\newcommand{\mathbfh}[1]{\hat{\mathbf{#1}}}

\newcommand{\mi}[0]{\mathrm{i}}

\begin{document}

\title{Nonequilibrium Thermodynamics of Quantum Friction}

\author{D. Reiche}
\email{reiche@physik.hu-berlin.de}
\affiliation{Humboldt-Universit\"at zu Berlin, Institut f\"ur Physik, AG
			Theoretische Optik \& Photonik, 12489 Berlin, Germany}
\affiliation{Max-Born-Institut, 12489 Berlin, Germany}

\author{F. Intravaia}
\affiliation{Humboldt-Universit\"at zu Berlin, Institut f\"ur Physik, AG
			Theoretische Optik \& Photonik, 12489 Berlin, Germany}

\author{J.-T. Hsiang}
\affiliation{Center for High Energy and High Field Physics, National Central University, Chungli 32001, Taiwan}

\author{K. Busch}
\affiliation{Humboldt-Universit\"at zu Berlin, Institut f\"ur Physik, AG
			Theoretische Optik \& Photonik, 12489 Berlin, Germany}
\affiliation{Max-Born-Institut, 12489 Berlin, Germany}

\author{B. L. Hu}
\affiliation{Joint Quantum Institute and Maryland Center for Fundamental Physics, University of Maryland, College Park, Maryland 20742-4111, USA}

\begin{abstract} 
Thermodynamic principles are often deceptively simple and yet surprisingly powerful.  
We show how a simple rule, such as the net flow of energy in and out of a moving atom 
under nonequilibrium steady state condition, can expose the shortcomings of many 
popular theories of quantum friction.  
Our thermodynamic approach provides a conceptual framework in guiding atom-optical 
experiments, thereby highlighting the importance of fluctuation-dissipation 
relations and long-time correlations between subsystems. 
Our results introduce consistency conditions for (numerical) models of nonequilibrium dynamics of open quantum systems.
\end{abstract}

\maketitle


\paragraph{Introduction.}
Fluctuations have a profound impact on physical reality, ranging from weak yet measurable 
forces all the way to structure formation in the universe. 
In the quantum realm, the existence of fluctuation-induced interactions was confirmed by pioneering 
\cite{derjaguin56,sukenik93} and ensuing experiments with increasing accuracy and scope
\cite{chan01,bimonte16,tang17,chan18,cunuder18,garrett18,haslinger18,norte18,sedmik18,somers18,peyrot19}. 

Many theoretical approaches have been designed to explain each distinct manifestation
of these quantum fluctuation phenomena.
However, a broader perspective is captured by the fluctuation-dissipation theorem (FDT):
For an open system in equilibrium, this theorem expresses the detailed balance between incoming and 
outgoing power, ensuring that the system is in a state of maximal entropy \cite{kubo66}. 
When nonequilibrium conditions prevail, the description of quantum 
fluctuation-induced phenomena is remarkably more involved and, to the best 
of our knowledge, general FDTs for fully nonequilibrium systems are lacking.  
Instead, a convenient assumption known as local thermal equilibrium (LTE) is often invoked \cite{polder71}.
This assumption significantly reduces the mathematical complexity of the problem and was broadly 
applied to the situation of temperature gradients between macroscopic bodies 
\cite{antezza05,benabdallah11,krueger11a,desutter19,maghrebi19}, 
atom-surface forces in thermal 
\cite{behunin11,dedkov17} 
as well as mechanical 
\cite{dedkov17,manjavacas17} 
nonequilibrium or under the influence of external driving fields 
\cite{rodriguezlopez17}, 
and for computing the radiation of a relativistic electron close to an interface 
\cite{rivera19}.
However, the theoretical basis for LTE and the conditions in which it fails to apply are usually 
not so well discussed:
First, under nonequilibrium conditions, the detailed balance (which is implicitly contained in LTE) 
is broken and, second, LTE is known to often disregard backaction of the environment 
\cite{grabert88,hu92,ford96}. 
In quantitative terms, LTE was already proven to be insufficient in the context of 
atom-surface quantum friction, e.g., underestimating the force by roughly half \cite{intravaia16a} or misrepresenting other important mechanisms \cite{intravaia19a}.  

In the framework of nonequilibrium atom-surface interactions, other often used methods have their 
own strengths and shortcomings. For instance, the Born-Markov approximation (BM) 
\cite{scheel09,Note1} or a perturbative treatment of the atomic level shift \cite{wylie85,svidzinsky19} do not rely on equilibrium. 
However, with regards to backaction and memory effects, these methods can only partially capture the impact of the environment 
\cite{intravaia14}. 
For quantum friction, they have been shown to lead to an incorrect velocity scaling 
\cite{intravaia16,klatt17} 
or erroneously predict exponentially vanishing forces (see the discussion in 
Refs.~\cite{intravaia14,dedkov17}).

In this Rapid Communication we address the deficiencies of these commonly used assumptions and 
approximations from another perspective, namely, the nonequilibrium thermodynamics 
of quantum friction. 
Even when the discrepancy between the approximate and the more carefully derived results 
might seem to be quantitatively marginal on the level of forces, the errors become manifest and 
easily identifiable when one applies the thermodynamic principles. 
In fact, neglecting the memory of the interaction or the long-time correlations between the system 
and environment -- as the BM and the LTE assumption do -- can lead to non-existent thermodynamic 
instabilities, such as, in the case of quantum friction, an over-time increase to infinity of 
the internal energy of the atom.
Our cure for this is the thermodynamic principle-enforced, self-consistent (backaction-including) treatment of the relevant nonequilbrium quantum processes.
This provides us with a benchmark to identify and explain why other approximate theories succeed or fail.


\paragraph{Physical model.}
We consider an atom moving at nonrelativistic velocity $v$ along an axis of translational symmetry relative to one or an entire arrangement of several macroscopic objects with an arbitrary cross-sectional shape. 
These objects comprise non-magnetic, reciprocal, and spatially homogeneous materials.
We assume that the atom moves at a distance from the objects which is much larger than its size. 
Within a multipolar approach \cite{power80,cohen-tannoudji89}, this allows us to
focus on the fluctuation-induced interaction between the atomic electric dipole moment $\mathbfh{d}$ and the material-modified fluctuating electric field $\mathbfh{E}$. 
We also demand that the atom's center of mass approximately obeys a classical trajectory. 
This implicitly includes the 
existence of an external ``agent'' driving the atom in such a way as to maintain uniform motion.
We assume that the backaction of our total system, composed of the atom+field+matter, on the agent 
is sufficiently small compared to the force the agent exerts on the system to keep the atom moving 
at uniform velocity. 
Thus we can safely consider the inflow of energy to the moving atom from the agent separately from
the outflow of energy from the atom to the field modified by the material. The backaction of the 
material-modified field on the atom appearing as quantum friction is of course included, it being 
the main character in the drama \cite{Note2}. 
Finally, we assume zero temperature and an initial state factorized in the distant past~\cite{Note3}.

In the static case ($v=0$), it can be shown that such a dynamical system equilibrates at late times 
\cite{hsiang18}. 
For atomic velocities $v\neq0$, however, the state of the system can deviate from the \textit{global} 
equilibrium condition \cite{deffner11}. Also, for finite coupling strength, the system and environment 
are inseparably intertwined and the assumption that equilibrium ensues \textit{locally} is not warranted.
Yet, dissipation (e.g. in the material) leads to finite correlation times between the system and environment establishing irreversibility in the interaction. 
When the different irreversible processes balance, the dynamics of the system becomes stationary and 
it reaches a nonequilibrium steady state (NESS) 
\cite{sasa06}. 
Such a state is thermodynamically characterized by the existence of a non-vanishing current of energy, sourced by some external drive \cite{seifert10} or temperature gradient \cite{lopez18,hsiang20a}, that compensates for
all different forms of losses in the system [Eq. \eqref{Eq:PRad}]. 
For the atomic subsystem the NESS requires a balance between incoming $P_{\rm in} $ 
and outgoing power $P_{\rm out}$ from and to the material-modified vacuum, respectively. 
If otherwise, the atomic energy would be increasing indefinitely contradicting the stationarity and the stability of the atomic dynamics. 
In the following, in lieu of a rigorous proof of the existence of the NESS 
\cite{Note4}, we provide an explicit late-time solution for a specific model [see Eq. \eqref{Eq:Langevin}] 
and show that the anticipated power balance $P_{\rm in}=P_{\rm out}$ holds, but only under certain 
conditions. 
This supplies a physical reasoning for its existence in more general contexts. 

Moving at constant velocity, the atom's internal degrees of freedom are in continuous 
exchange of energy, translational and angular momentum with the surrounding material-modified 
quantum field. 
In the steady-state, these processes can be described in 
terms of the three-dimensional Langevin equation \cite{Note5}
\begin{align}\label{Eq:Langevin}
	\frac{\ddot{\mathbfh{d}}(t)
	+\omega_a^2\mathbfh{d}(t)}
	{\alpha_0\omega_a^2}
	+2\int_0^{\infty}\mathrm{d}\tau~
	\underline{\gamma}(\tau,v)\cdot
	\dot{\mathbfh{d}}(t-\tau)
	=\hat{\bm{\xi}}(t,v),
\end{align}
where $\alpha_0$ is the atomic static polarizability and $\omega_a$ the bare resonance frequency of the lowest energetically accessible dipole transition.
Here, $\alpha_0$ plays the role of the coupling constant between the microscopic object and its surrounding electromagnetic environment.
The quantum Langevin force and the dissipative memory kernel, respectively, can be written as
 \cite{intravaia19a}
\begin{subequations}
\label{Eq:LangFMemoryK}
\begin{align}
\hat{\bm{\xi}}(t,v)
	&=
	\int\frac{\mathrm{d}\omega}{2\pi}\int\frac{\mathrm{d} q}{2\pi}
	\mathbfh{E}_{0}(q,\mathbf{R}_a,\omega)
	e^{-\mathrm{i}\omega_{q}^{-}t},
	\end{align}
	\begin{align}
	\underline{\gamma}(t,v)
	&=
	\int\frac{\mathrm{d}\omega}{2\pi}	
	\int\frac{\mathrm{d} q}{2\pi}
	~
	\frac{\underline{G}_{\Im}(q,\mathbf{R}_a,\omega)}{\omega_{q}^{-}}
	e^{-\mathrm{i}\omega_{q}^{-} t}.
\end{align}
\end{subequations}
Here, $q$ is the component of the radiation's wave vector in the direction of motion, while 
$\mathbf{R}_a$ is the atom's position in the plane orthogonal to it. We also defined the 
Doppler-shifted frequency as $\omega_{q}^{\pm}=\omega\pm q v$.  
The atomic system is driven by the fluctuations of the field in the absence of the atom,
$\mathbfh{E}_{0}$. 
The dispersion as well as dissipation mechanisms are encoded in the Green's tensor $\underline{G}$ 
with $\underline{G}_{\Im}=(\underline{G}-\underline{G}^{\dagger})/(2\mathrm{i})$.
$\underline{G}$ solves the Maxwell equations with appropriate boundary conditions 
and hence incorporates the material properties, the translational symmetry of our system, and 
ensures the causality of the interaction \cite{tomas95}.
Consequently, $\underline{G}_{\Im}$ is a Hermitian positive semidefinite matrix 
for $\omega>0$, while a stationary and a causal dynamics of the dipole implies that 
$\underline{\gamma}(\omega,v)$ must be positive definite.
Since without the moving atom the system is in equilibrium, the field $\mathbfh{E}_{0}$ 
must satisfy the FDT
\begin{align}\label{Eq:FDTVac}
&\langle
\mathbfh{E}_{0}(q,\omega)
\mathbfh{E}_{0}(q',\omega')
\rangle
	\\\nonumber
	&\qquad=
	\hbar(2\pi)^2\text{sgn}(\omega)
	\underline{G}_{\Im}(q,\mathbf{R}_a,\omega)
	\delta(\omega+\omega')
	\delta(q+q'),
\end{align}
where $\text{sgn}(\omega)$ is the sign function and $\delta(x)$ the Dirac delta. 
Hereafter we consider the symmetric quantum average, i.e., $\langle\hat{A}\hat{B}\rangle\equiv\langle\hat{A}\hat{B}+\hat{B}\hat{A}\rangle/2$ \cite{FootnoteSymmetricAverage,dalibard82}.
Equation \eqref{Eq:Langevin} is solved in the Fourier domain as $\mathbfh{d}(\omega,v)=\underline{\alpha}(\omega,v)\cdot\hat{\bm{\xi}}(\omega,v)$ by means of the dressed and velocity-dependent atomic polarizability $\underline{\alpha}(\omega,v)$ (see Ref.~\cite{SuppMat} for details). 
Physically, the latter contains spontaneous emission \cite{intravaia16b}, dispersion and dissipation due to the presence of the material \cite{reiche17}.
The correlation matrix of the Langevin force becomes stationary and real in the steady state \cite{FootnoteSymmetricAverage}, i.e., $\langle\hat{\bm{\xi}}(t,v)\hat{\bm{\xi}}(t',v)\rangle\equiv\hbar\, \underline{\nu}(t,t',v)\to\hbar\,\underline{\nu}(\tau,v)$ ($\tau\equiv t-t'$). 
Moreover, the quantum noise is colored:
\begin{align}\label{Eq:Nu}
\underline{\nu}(\omega,v)
	&=
	\int\frac{\mathrm{d} q}{2\pi}~
	\text{sgn}(\omega_{q}^{+})
	\underline{G}_{\Im}(q,\mathbf{R}_a,\omega_{q}^{+}).
\end{align}
Our self-consistent treatment of the system [Eq. \eqref{Eq:Langevin}] describes the connection between field fluctuations and dipole fluctuations via the relation
\begin{align}\label{Eq:PowerSpectrum}
\langle
\mathbfh{d}(\omega)\mathbfh{d}(\omega')
\rangle
	=2\pi \hbar\, \underline{\Sigma}(\omega,v)
	\,\delta(\omega+\omega')
	,
\end{align}
where $\underline{\Sigma}(\omega,v)=\underline{\alpha}(\omega,v)\underline{\nu}(\omega,v)\underline{\alpha}^{\dag}(\omega,v)$ is positive semidefinite for all $\omega$ because of the properties of all involved matrices \cite{SuppMat}.
The relations in Eqs. \eqref{Eq:Nu} and \eqref{Eq:PowerSpectrum} generalize the FDT to the NESS and lead to previously reported results on quantum friction \cite{intravaia16a,intravaia19a,reiche20d}. 
%


\paragraph{Nonequilibrium thermodynamics.}

We now examine the thermodynamic implications of Eqs. (\ref{Eq:Nu}) and (\ref{Eq:PowerSpectrum}).
The ``in'' and the ``out'' parts of the moving atom's energy flow per unit time are \cite{li93,hsiang15}
\begin{subequations}
\begin{gather}
P_{\mathrm{in}}=\langle\hat{\bm{\xi}}(t,v)\cdot\dot{\mathbfh{d}}(t)\rangle,\\
P_{\mathrm{out}}=2\int_0^{\infty}\mathrm{d}\tau~
	\langle\dot{\mathbfh{d}}(t)\cdot
	\underline{\gamma}(\tau,v)\cdot
	\dot{\mathbfh{d}}(t-\tau)\rangle,
\end{gather}
\end{subequations}
which yield a change in energy $E$ of the atom given by
$\dot{E}=P\equiv P_{\rm in}-P_{\rm out}$.
Using Eq. \eqref{Eq:Langevin}, we can show that $P=0$ in the NESS (see Ref.~\cite{SuppMat}), i.e., 
there is no net energy flow in or out of the system
since
\begin{equation}
P_{\rm in}= P_{\rm out}=2
	\int_0^{\infty}\frac{\mathrm{d}\omega}{2\pi}
	\hbar\omega
	~
	\text{Tr}
	\left[	
	\underline{\nu}(\omega,v)
	\underline{\alpha}_{\Im}(\omega,v)
	\right],
\end{equation}
where, similarly to $\underline{G}_{\Im}$, we defined $\underline{\alpha}_{\Im}=(\underline{\alpha}-\underline{\alpha}^{\dagger})/(2\mathrm{i})$ and ``$\text{Tr}$'' takes the trace of the resulting matrix.

A few comments are in order. 
First, $P_{\rm in/out}$ is positive since $\underline{\alpha}_{\Im}(\omega,v)$ 
is positive definite for $\omega\ge 0$ \cite{SuppMat}.
Notably, within our initial assumptions, the previous results hold for any (nonrelativistic) velocity 
and arbitrary functional frequency behavior of the memory kernel. 
In particular, the damping $\underline{\gamma}$ need not be Ohmic and it can contain any 
physical resonance of the system.

Second, a vanishing power is equivalent to the condition $\langle\dot{\mathbfh{d}}\cdot \mathbfh{E}\rangle=0$ 
in the NESS, where $\mathbfh{E}$ is the \textit{total} field acting on the moving dipole. 
This allows us to formulate a relation between the (mechanical) frictional force $F_{\rm fric}$ 
and the total power radiated from the particle into the environment $P_{\rm rad}$ \cite{intravaia15a,SuppMat}. 
We have $P_{\rm rad}=P_{\rm ext}\equiv-vF_{\rm fric}$, where
\begin{equation}\label{Eq:PRad}
P_{\rm rad}
	=
	2 
	\mathrm{Tr}
	\int_{0}^{\infty}
	\mathrm{d}\omega
	\int\frac{\mathrm{d}q}{2\pi} \omega\,
	\underline{S}^{\sf T}(-\omega_{q}^{-},v)
	\underline{G}_{\Im}(q,\mathbf{R}_{a},\omega),
\end{equation}
with ``${\sf T}$'' the transpose of a matrix.
Here, $\underline{S}$ is the atomic power spectrum tensor defined in previous work 
and for our system it has a form very similar to $\underline{\Sigma}$ \cite{intravaia16,SuppMat}. 
The expression for $F_{\rm fric}$ is instead recovered by replacing $\omega\to q$ in the previous integrand \cite{SuppMat}. 
The identity $P_{\rm rad}= P_{\rm ext}$ is the \emph{black-box approach} counterpart to the microscopic perspective offered by $P_{\rm out}=P_{\rm in}$. Although physically equivalent, $P_{\rm rad}= P_{\rm ext}$ provides a look from the outside of the microscopic object without paying attention to its internal dynamics.
It sets the accent on the balance between the total mechanical power entering the system (performed by the external agent balancing the frictional force)
and what is coming out as electromagnetic energy dissipated in the environment. 
Since it does not require a specific microscopic model for 
the atom's internal degrees of freedom \cite{intravaia16,SuppMat}, the relation between $P_{\rm rad}$ and $F_{\rm fric}$ offers therefore an alternative, more general 
perspective on the \textit{irreversible flow} of energy (accompanied by the production 
of entropy) through the system~\cite{horowitz20}.

\begin{figure}[t]
\begin{center}
  \includegraphics[width=0.48\textwidth]{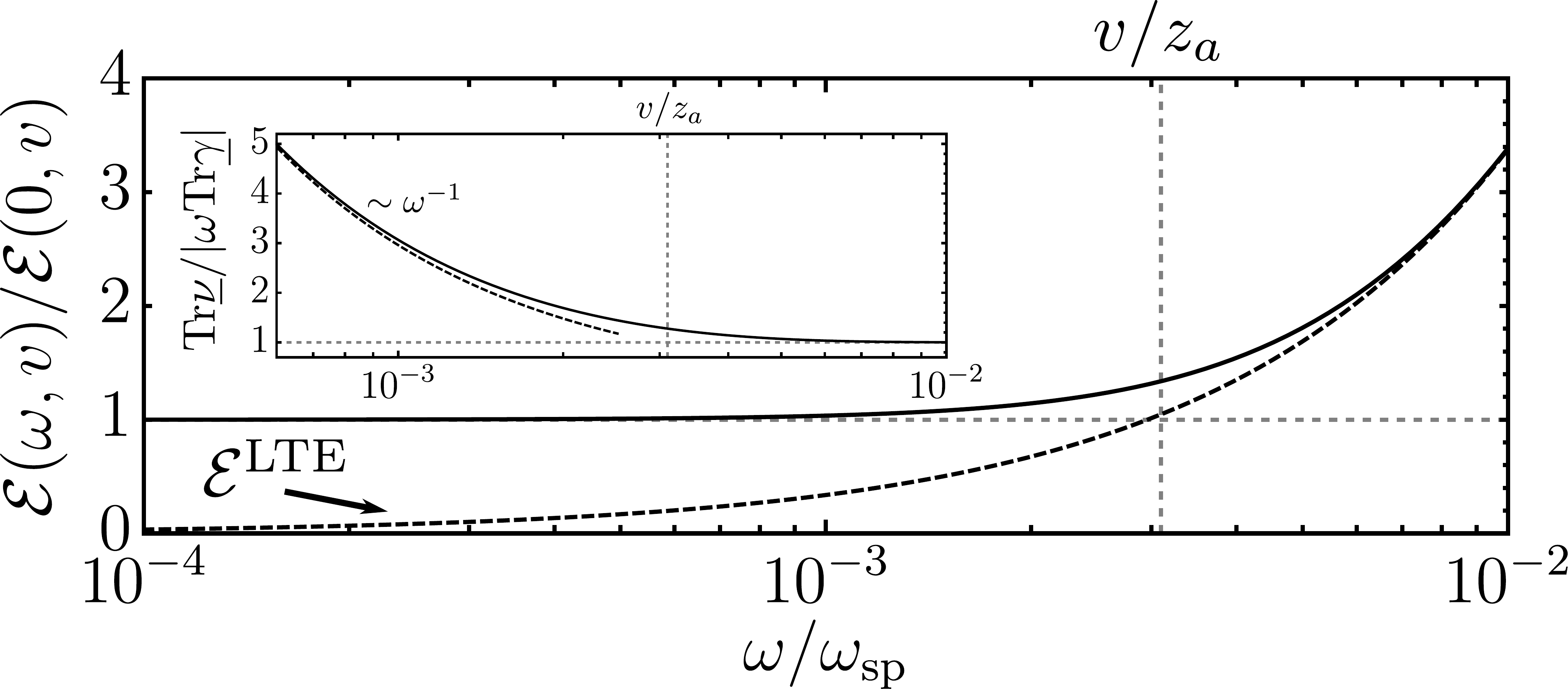}
\vspace{-.7cm}  
  \caption{Spectral energy for an atom moving parallel to a planar interface (solid line) and respective LTE result (dashed). 
  We employ the Drude model, where
  $
  r^{\rm TM}= \omega_{\rm sp}^2
  [\omega_{\rm sp}^2-\omega^2-\mi\Gamma\omega]^{-1}
  $ 
  \cite{SuppMat} with $\omega_{\rm sp}$ the surface plasmon-polariton resonance and $\Gamma$ 
	the associated damping.
  We set $v=10^{-4}c$, $z_a=1$\,nm, and use parameters for gold \cite{barchiesi14}.
  Inset: 
  Fluctuation-dissipation inequality and the asymptote of Eq.~\eqref{Eq:FDTInEqAsymp} (dashed).
  \label{Fig:Inequality}  }
   \end{center}
   \vspace{-0.7cm}  
\end{figure}

Third, $P=0$ implies that the total energy $E$ corresponding to the atom's internal dynamics is constant 
\cite{Note6}.
From Eqs.~\eqref{Eq:Langevin}-\eqref{Eq:PowerSpectrum}, $E$ can be written as an integral over positive 
frequencies of the spectral density \cite{hsiang18,SuppMat}
\begin{align}
\label{Eq:SpectralEnergy}
\mathcal{E}(\omega,v)
	&=\frac{\hbar}{2\pi}
	\frac{\omega_a^2+\omega^2}{\omega_a^2}
	\text{Tr}
	\left[\frac{\underline{\Sigma}(\omega,v)
	}{\alpha_0}
	\right]\ge 0.
\end{align}
Typically, $\alpha_{0}/\epsilon_0\ll \lambda^{3}$ (weak coupling), where $\lambda$ is a length scale which characterizes the system's behavior: It is connected to the system's specific properties (e.g., the optical response of the involved objects as well as their positions and geometries) through the electromagnetic Green's tensor.
In this weak-coupling limit and in equilibrium ($v=0$), we have $E\to 3\hbar\omega_a/2$ as 
expected, while a stronger coupling would effectively modulate the value of the atomic energy \cite{hsiang18,li93}.
Deviating from equilibrium ($v\neq0$), the energy becomes an even function of the velocity 
and at the leading order in $\tilde{\alpha}_0\equiv\alpha_{0}/(\epsilon_0\lambda^{3})$ we have
\begin{equation}
\label{lowFrequenciesE}
\mathcal{E}(0,v)
	\propto
	\tilde{\alpha}_0\,\varepsilon(v)\neq 0,
\end{equation}\noindent where $\varepsilon$ is a function of velocity with $\varepsilon(0)= 0$ \cite{SuppMat}. 
Equation~\eqref{lowFrequenciesE} is thermodynamically related to 
the stationary energy flow through the atom in the NESS and highlights two important aspects of our 
analysis: On the one hand, low frequencies (long-time correlations) play an important role in correctly 
capturing the nonequilibrium physics of the system. 
On the other hand, in equilibrium, $\mathcal{E}$ vanishes for $\omega\to 0$, in agreement with the FDT 
and with a thermodynamically consistent description of a dissipative atomic system at $T=0$.
In contrast, assuming local equilibrium enforces $\mathcal{E}(0,v)=0$ for all atomic velocities (see 
also Fig.~\ref{Fig:Inequality}). 
Similarly, within the BM or a related perturbative treatment, even at $v=0$, $\mathcal{E}$ approaches a nonzero 
constant for $\omega\to0$, whose value depends on the involved dissipative mechanisms and might be related to the initial state preparation \cite{intravaia16,SuppMat}. 
This means then that, in different ways, 
both the LTE and the BM descriptions misrepresent the low frequency contributions to the system's 
dynamics.
Specifically for our system, Eqs.~\eqref{Eq:SpectralEnergy}, \eqref{lowFrequenciesE} and the expressions 
for $F_{\rm fric}$ \cite{SuppMat} imply that an adequate description of the nonequilibrium process 
requires at least $O(\tilde{\alpha}_0^2)$.
Consequently, the thermodynamical consistency and/or the accuracy of results that address the frictional process to first order in the atomic polarizability can be questionable and must be interpreted 
with care, depending on the specific approach being employed as well as on the dissipative mechanisms at work in the system. 
For instance, previous work has shown that, although the LTE assumption for quantum friction can be 
justifiable to some extent at orders $O(\tilde{\alpha}_0)$ for a particle dynamics that allows 
for strong \emph{intrinsic} dissipation (e.g. for metallic nanoparticles), it fails when 
radiation-induced damping prevails and backaction is relevant \cite{intravaia16a,reiche20d}.

Finally, it is important to underline that despite its direct appeal, the result $P=0$ 
is technically non-trivial to realize. 
It could only be achieved with careful ``bookkeeping'' of the system's full roto-translational spectrum of correlations taking the backaction from the environment fully into account 
[Eq. \eqref{Eq:PowerSpectrum}]. 
Any deviation from this complete self-consistency can lead to thermodynamical instabilities. 
This is indeed the case for the LTE approach, which amounts to replacing
$\underline{\nu}(\omega,v)
\to
\omega\text{sgn}(\omega)
\underline{\gamma}(\omega,v)$
in Eq. (\ref{Eq:PowerSpectrum}). 
It effectively neglects the Doppler-shift of the radiation in the evaluation of the sign-function 
in $\underline{\nu}(\omega,v)$ and breaks the total power balance, contradicting the stationarity 
condition for NESS. 
In this case we have \cite{SuppMat}
\begin{align}
\label{powerLTE}
P^{\mathrm{LTE}}
&=2\int_0^{\infty}\frac{\mathrm{d}\omega}{2\pi}
	\hbar\omega \mathrm{Tr}\left[\left\{\underline{\nu}(\omega,v)-\omega \underline{\gamma}
	(\omega,v)\right\}
	\underline{\alpha}_{\Im}(\omega,v)
	\right]
	\nonumber\\
&\equiv P^{\mathrm{LTE}}_{\rm in}-P^{\mathrm{LTE}}_{\rm out}\not=0.	
\end{align}
This is the thermodynamic evidence that not including nonequilibrium backaction in perturbative 
approaches or simplifying assumptions can lead to glaring mistakes. 
In contrast, nonequilibrium dynamics with self-consistent backaction is fully guaranteed from 
the thermodynamic principles which we invoke.
\paragraph{Fluctuation-dissipation inequality.} 
Equation~\eqref{powerLTE} shows that the relation between the quantum fluctuations
$\underline{\nu}(\omega,v)$ and the dissipative memory kernel $\underline{\gamma}(\omega,v)$ gives a measure of the impact of nonequilibrium onto the system. 
If we define $\underline{\tilde G}_{\Im}(q,\mathbf{R}_{a},\omega)=\text{sgn}(\omega)\underline{G}_{\Im}(q,\mathbf{R}_{a},\omega)$ and use the identity 
$\text{sgn}(x)[\text{sgn}(x)\pm 1]=2\theta(\pm x)$,
we can write
\begin{equation}
	\label{inequality1}
	\underline{\nu}(\omega,v)\pm\omega\underline{\gamma}(\omega,v)=\int\frac{\mathrm{d}q}{\pi}\,
	\theta(\pm\omega_{q}^{+})
	\underline{\tilde G}_{\Im}(q,\mathbf{R}_{a},\omega_{q}^{+}),
\end{equation}
which is Hermitian and positive semidefinite for all values of $q$ and $\omega$.
We can then conclude that for our system $P^{\mathrm{LTE}}\ge 0$ for all velocities 
and colors of the noise.
Also, using the Loewner order \cite{loewner34}, in accordance with the fluctuation-dissipation 
inequality put forward in Ref.~\cite{fleming13}, we can write $\underline{\nu}(\omega,v)\ge|\omega\underline{\gamma}(\omega,v)|$. 
Specifically, this indicates that in the NESS the field's fluctuations ($\underline{\nu}$) are always equal to or exceed the field's induced dissipative power ($\omega\underline{\gamma}$) \cite{fleming13}. 
The matrix $\underline{\nu}(\omega,v)-\omega \underline{\gamma} (\omega,v)$ only goes to zero 
either for $v=0$ restoring the equilibrium FDT, or asymptotically for frequencies $\omega\gg v/\lambda$.
In agreement with the behavior of the energy spectral density [Eq. \eqref{lowFrequenciesE}], 
the largest deviations occur at low frequencies ($\omega\ll v/\lambda$), emphasizing once again the connection of these low frequencies to the nonequilibrium dynamics of our system. 
Physically, this shows that
simply using the equilibrium FDT neglects the interaction energy that corresponds to correlation 
times larger than $\lambda/v$ (of the order of nanoseconds for typical values). 
These correlations are an inalienable part of the system interacting with its environment and an 
important feature of nonequilibrium settings.
The fluctuation-dissipation inequality quantifies this mismatch and the complete description of 
the system requires a more careful treatment by means of the generalized FDT [Eq. \eqref{Eq:Nu}].

%

To obtain quantitative insight, it is interesting to consider the case of an atom moving at 
a distance $z_{a}\sim\lambda$ close to a planar interface separating vacuum from an infinite 
half space composed of a typical Ohmic dissipative and spatially local material (Fig.~\ref{Fig:Inequality}) 
\cite{Note7}. 
For this geometry, the analytic expression for the Green's tensor is known \cite{tomas95}. 
Since $v/z_{a}$ is usually in the material's Ohmic region, we can write \cite{SuppMat}
\begin{align}
\label{Eq:FDTInEqAsymp}
\frac{\mathrm{Tr}[\underline{\nu}(\omega,v)]}
	 {\left|
	\omega\mathrm{Tr}[\underline{\gamma}(\omega,v)]
	\right|}=
\begin{cases}
	1, & \omega\gg \frac{v}{z_{a}}, \\
	\frac{3}{\pi\omega}~
	\frac{v}{z_a}~
	&\omega\ll \frac{v}{z_{a}}.
\end{cases}			
\end{align}
Equation \eqref{Eq:FDTInEqAsymp} shows that the usual FDT holds for $v=0$. 
However, at nonzero velocity, it prescribes a finite low-frequency domain encoding 
corrections to the nonequilibrium statistics of the system.
For the same setup, at the leading order in $\alpha_0$ and $v$, the net power within 
the LTE approach evaluates to \cite{SuppMat,dedkov17}
\begin{align}\label{Eq:PLTE}
P^{\rm LTE}
	&\sim
	\hbar
	\frac{45}{4}
	\frac{v^4}{(2\pi)^3}
	\frac{\alpha_0^2}{\epsilon_0^2}
	\frac{\text{Im}\left\{\lim_{\omega\to 0}\partial_{\omega}r^{\rm TM}\right\}^{2}}{(2z_a)^{10}}\ge 0,
\end{align} 
where $\epsilon_0$ is the vacuum permittivity and $r^{\rm TM}$ the bulk's transverse 
magnetic reflection coefficient.
As expected, $P^{\rm LTE}$ is positive for $v\neq0$, showing the LTE to fail at 
$O(\tilde{\alpha}_0^2)$ \cite{intravaia15a,SuppMat}.



%
\paragraph{Conclusions.}
The existence of a nonequilibrium steady state in a dissipative open quantum system 
implies the balance of energy flow in and out of the system. 
Our analysis shows that this condition imposes strict constraints on how different 
contributing factors should behave to meet the stringent self-consistency requirements in how the system interacts 
with its environment and how the latter back-acts on the system.  
Our formalism is rather general, it does not rely on a transient behavior, and can be readily applied to explore 
different materials and geometries with at least one direction of translational invariance.  
In addition, the full breadth of our analysis transcends a specific context and 
similar arguments can be made for other phenomena such as heat transfer \cite{polder71,joulain05,hsiang15,barton16}.

The physical consistency condition which underlies our results can also serve even
broader purposes. 
With increasing computational power, there has been a surge of interest in the field of 
photonics in \textit{design and inverse design}, where one aims to find suitable physical 
setups for given functional characteristics using numerical optimization procedures 
\cite{molesky18}. 
In nonequilibrium setups, this is a particularly complicated problem since one is mostly 
concerned with vector-valued quantities and a complex resonance structure that can lead 
to numerical obstacles 
\cite{reid17}. 
Also, due to the lack of analytical solutions, one has to rely on limiting scenarios as well as 
more general properties based on the system's symmetries for validating the obtained result.
Power balance and the described inequality hence serve as a benchmark for such nonequilibrium 
calculations. 
Additionally, due to the extensive efforts in controlling atomic systems (see also Refs. 
\cite{sague07,elliott18,becker18,barrett19} in addition to the above), the principles and 
methodology presented here can be used for experimentally understanding and probing nonequilibrium fluctuation theorems \cite{horowitz20} and entropy production in 
nonequilibrium situations \cite{deffner11,ness17}. 
In particular, this means that we can provide a general proof of what is often found case by case 
based on partially justifiable assumptions.
Experimentally, when signatures of quantum friction are detected, our criteria can be used 
to ascertain and discriminate whether it truly originates from nonequilibrium quantum 
fluctuations.

%

\paragraph{Acknowledgments.}
We thank R. O. Behunin, K. Sinha, B. Beverungen and D.-N. Hyunh as well as an anonymous referee for interesting discussions. 
We acknowledge support from the Deutsche Forschungsgemeinschaft (DFG, German Research 
Foundation) -- Project-ID 182087777 -- SFB 951.
D.R. is grateful for support from the German-American Fulbright Commission (Doktorandenprogramm) 
and thanks the Maryland Center for Fundamental Physics and the Joint Quantum Institute for 
hospitality.
K.B. thanks F. Herrmann and the late G. Falk for introducing him to irreversible thermodynamics.

\appendix




\begin{thebibliography}{10}
\bibitem{FootnoteSymmetricAverage}
In the case of vector operators, the symmetric average has to be understood component-wise, i.e., $\langle\mathbfh{A}\mathbfh{B}\rangle_{ij}\equiv\langle \hat{A}_i\hat{B}_j\rangle\equiv\langle \hat{A}_i\hat{B}_j+\hat{B}_j\hat{A}_i\rangle/2$. 
Using dyadic products, this can be written as
$
\langle\mathbfh{A}\mathbfh{B}\rangle\equiv [\langle\mathbfh{A}\mathbfh{B}\rangle+\langle\mathbfh{B}\mathbfh{A}\rangle^{\sf T}]/2.
$
The symmetric ordering renders real all quantum averages over products of Hermitian operators \cite{dalibard82}.
\end{thebibliography}
%


\clearpage
\setcounter{page}{1}
\setcounter{figure}{0}
\setcounter{equation}{0}
\renewcommand{\theequation}{S\arabic{equation}}
\renewcommand{\figurename}{\textbf{Supplementary Figure}}
\renewcommand{\thefigure}{{\bf S\arabic{figure}}}

\section*{\Large Supplemental Material}

\section{The nonequilibrium power flux}
The 3D-quantum Langevin equation [Eq.~(1) of the main text], describing the atom's internal dynamics, is solved by $\mathbfh{d}(\omega,v)=\underline{\alpha}(\omega,v)\cdot\hat{\bm{\xi}}(\omega,v)$ (stationary solution), where 
\begin{align} \label{Eq:Pol}
\underline{\alpha}(\omega,v)
	&=
	\alpha_B(\omega)
	\left[
	1-\alpha_B(\omega)\underline{\Delta}(\omega,v)
	\right]^{-1}
\end{align}
is the dressed velocity-dependent polarizability. The scalar function
$\alpha_B(\omega)=\alpha_0\omega_a^2/(\omega_a^2-\omega^2)$ is the atomic bare polarizability and, using the Kramers-Kronig relations, we have defined 
\begin{align}
\underline{\Delta}(\omega,v)
	&=
	\mathcal{P}
	\int\frac{\mathrm{d}\bar{\omega}}{\pi}
	\frac{\bar{\omega}\underline{\gamma}(\bar{\omega},v)}{\bar{\omega}-\omega}
	+\mathrm{i}\omega\underline{\gamma}(\omega,v)
	\nonumber\\
	&=
	\int\frac{\mathrm{d} q}{2\pi}
	\underline{G}(q,\mathbf{R}_a,\omega_{q}^{+})
\end{align}
with $\mathcal{P}$ the Cauchy principal value.
The integration goes over the whole real axis if not indicted otherwise. 
The properties of the Green tensor yield some important relations:
$\underline{G}(-q,\mathbf{R}_{a},  \omega)=\underline{G}^{\sf T}(q,\mathbf{R}_{a},\omega)$ and 
$\underline{G}^{*}(q,\mathbf{R}_{a},  \omega)=\underline{G}(-q,\mathbf{R}_{a}, -\omega)$.
They imply that 
$\underline{\alpha}(-\omega,v)=\underline{\alpha}^{*}(\omega,v)$ and $\underline{\alpha}^{\dag},(\omega,v)=\underline{\alpha}(-\omega,-v)$
as well as the identity
\begin{align}
\underline{\alpha}_{\Im}(\omega,v)
	&=
	\int\frac{\mathrm{d}q}{2\pi}
	\underline{\alpha}(\omega,v)\underline{G}_{\Im}(q,\mathbf{R}_a,\omega_{q}^{+})
	\underline{\alpha}^{\dagger}(\omega,v)
	\nonumber\\
	&=\omega
	\underline{\alpha}(\omega,v)\underline{\gamma}(\omega,v)
	\underline{\alpha}^{\dagger}(\omega,v),
\end{align}
where, in analogy to $\underline{G}_{\Im}$, we defined $\underline{\alpha}_{\Im}=(\underline{\alpha}-\underline{\alpha}^{\dagger})/(2\mathrm{i})$. 
Since $\underline{\gamma}$ is positive semidefinite so is $\underline{\alpha}_{\Im}$ for $\omega\ge 0$.
Our self-consistent description also leads to the matrix $\underline{\nu}$ and to the definition of $\underline{\Sigma}$ which are also positive semidefinite (see its definition and the discussion around Eq.~(5) of the main text).
 
From the previous expressions, we can write the power flowing into the atomic subsystem due to fluctuations, $P_{\mathrm{in}}=
	\langle\hat{\bm{\xi}}(t,v)\cdot\dot{\mathbfh{d}}(t)\rangle$, as follows
\begin{align}
P_{\mathrm{in}}
	&
	=
	\int\frac{\mathrm{d}\omega}{2\pi}
	\int\frac{\mathrm{d}\omega'}{2\pi}
	~
	(-\mi\omega')
	e^{-\mi(\omega+\omega')t}
	\nonumber\\
	&\qquad\times \mathrm{Tr} \left[\underline{\alpha}^{\sf T}(\omega',v)
	\langle
	\hat{\bm{\xi}}(\omega,v)
	\hat{\bm{\xi}}(\omega',v)
	\rangle\right]
	\nonumber\\
	&
	=
	\int\frac{\mathrm{d}\omega}{2\pi}
	~
	\mi \hbar\omega
	 \mathrm{Tr} \left[\underline{\alpha}^{\dag}(\omega,v)
	\nu(\omega,v)
	\right]
	\nonumber\\
	&=2
	\int_0^{\infty}\frac{\mathrm{d}\omega}{2\pi}
	\hbar\omega
	~
	\mathrm{Tr}
	\left[	
	\underline{\nu}(\omega,v)
	\underline{\alpha}_{\Im}(\omega,v)
	\right],
\end{align}
where we used that $\mathrm{Tr}[\underline{A}^{\sf T}]=\mathrm{Tr}[\underline{A}]$ for any matrix $\underline{A}$ and $\underline{\nu}(-\omega,v)=\underline{\nu}^{\sf T}(\omega,v)$.
Similarly, the power leaving the atomic subsystem due to dissipation reads
\begin{align}
P_{\mathrm{out}}
	&=
	2
	\int_0^{\infty}
	\mathrm{d}\tau~ \mathrm{Tr}\left[
	\underline{\gamma}^{\sf T}(\tau,v)
	\langle
	\dot{\mathbfh{d}}(t)
	\dot{\mathbfh{d}}(t-\tau)
	\rangle\right]
	\nonumber\\
	&=
	2
	\int_0^{\infty}\frac{\mathrm{d}\omega}{2\pi}
	~
	\hbar\omega
	\int\frac{\mathrm{d} q}{2\pi}
	\nonumber\\
	&\quad\times
	\mathrm{Tr}\left[
	\underline{G}_{\Im}(q,\mathbf{R}_a,\omega_{q}^{+})
	\underline{\alpha}(\omega,v)
	\underline{\nu}(\omega,v)
	\underline{\alpha}^{\dag}(\omega,v)
	\right]
	\nonumber\\
	&=
	2
	\int_0^{\infty}\frac{\mathrm{d}\omega}{2\pi}
	~
	\hbar\omega
	\int\frac{\mathrm{d} q}{2\pi}
	\nonumber\\
	&\quad\times
	\mathrm{Tr}\left[
	\underline{\alpha}(-\omega,-v)
	\underline{G}_{\Im}(q,\mathbf{R}_a,\omega_{q}^{+})
	\underline{\alpha}^{\dag}(-\omega,-v)
	\underline{\nu}(\omega,v)
	\right]
	\nonumber\\
	&=
	2
	\int_0^{\infty}\frac{\mathrm{d}\omega}{2\pi}
	~
	(-\hbar\omega)
	\mathrm{Tr}\left[
		\underline{\nu}(\omega,v)
	\underline{\alpha}_{\Im}(-\omega,-v)
	\right].
\end{align}
Since $\underline{\alpha}_{\Im}(-\omega,-v)=-\underline{\alpha}_{\Im}(\omega,v)$ it follows that
\begin{equation}
P_{\rm in}=P_{\rm out} \Rightarrow P=P_{\rm in}-P_{\rm out}=0 \quad\forall v.
\end{equation}
The self-consistency of our treatment is central for obtaining the previous result. A deviation from it can lead to a steady nonzero power transfer to the atomic system.
Even if this value is small, it will be accumulating over time and the consequences of making artificial assumptions on the underlying statistics of the interaction can be dramatic.
An example is the impact of the local thermal equilibrium (LTE) assumption.
This approach does not modify $\langle\hat{\bm{\xi}}(\omega,v)\hat{\bm{\xi}}(\omega',v)\rangle$
effectively leading to
\begin{align}
P_{\mathrm{in}}=P_{\mathrm{in}}^{\mathrm{LTE}}.
\end{align}
Instead, $P_{\mathrm{out}}$ is intimately related to the nonequilibrium relation in Eq.~(5). 
The LTE approach assumes the dipole correlations to fulfill the equilibrium FDT leading to
\begin{align}
\label{dipoleLTE}
\langle
\mathbfh{d}(\omega)\mathbfh{d}(\omega')
\rangle
	\stackrel{\rm LTE}{=}
	2\pi\,
	\hbar\,\text{sgn}(\omega)\,\underline{\alpha}_{\Im}(\omega,v)\,
	\delta(\omega+\omega').
\end{align}
In some cases even the dependence of the polarizability on the velocity is 
ignored \cite{dedkov17}.
The LTE assumption then modifies the outgoing power as follows
\begin{align}
P_{\mathrm{out}}^{\mathrm{LTE}}
		&=
	2
	\int_0^{\infty}\frac{\mathrm{d}\omega}{2\pi}
	~
	\hbar\omega
	\mathrm{Tr}\left[\text{sgn}(\omega)\,\omega \underline{\gamma}(\omega,v)
	\underline{\alpha}_{\Im}(\omega,v)
	\right]
	\nonumber\\
	&\neq P_{\mathrm{out}}	.
\end{align}
Equation \eqref{dipoleLTE} eventually leads to an imbalance of the total power $P^{\mathrm{LTE}}=P_{\mathrm{in}}^{\mathrm{LTE}}-P_{\mathrm{out}}^{\mathrm{LTE}}$, i.e.
\begin{align}
P^{\mathrm{LTE}}
&=2\int_0^{\infty}\frac{\mathrm{d}\omega}{2\pi}
	~
	\hbar\omega
	\\\nonumber
	&\quad\times
	\mathrm{Tr}\left[\left\{\underline{\nu}(\omega,v)-\text{sgn}(\omega)\,\omega \underline{\gamma}(\omega,v)\right\}
	\underline{\alpha}_{\Im}(\omega,v)
	\right]
\end{align}
which is in general positive. 
Indeed, using the fluctuation-dissipation inequality (see the main text), the previous integrand contains the trace of two positive semidefinite matrices, which is always positive or zero. 

At low velocity and at the leading order in $\alpha_0$, the previous expression takes the form
\begin{align}
\label{lowvPLTE}
P^{\mathrm{LTE}}
	&\sim
	\frac{\hbar}{\pi}
	\frac{\alpha_0^2v^4}{12}
	\int\frac{\mathrm{d}q}{2\pi}
	\int\frac{\mathrm{d}q'}{2\pi}
	~
	q^2
	\left(
	q^2-2q q'
	\right)
	\nonumber\\
	&\times
	\mathrm{Tr}
	\left[
	\underline{G}_{\Im}'(q,\mathbf{R}_a,0)
	\cdot
	\underline{G}_{\Im}'(q',\mathbf{R}_a,0)
	\right],
\end{align}
where the prime indicates a derivative with respect to frequency.

Consider now the specific case of an atom moving along the $x$-axis, in front of planar metallic interface. 
For atom-surface separations $z_{a}$ smaller than the metal's plasma wavelength (typically $\sim 100$ nm or larger \cite{barchiesi14}), the Green tensor is dominated by its scattered part evaluated in the near-field limit. 
For a plane we have
\begin{align}\label{Eq:Green}
\underline{G}_{\Im}(q,\mathbf{R}_a,\omega)&\equiv \underline{G}(p_{x},z_a,\omega)
\nonumber\\
	&\sim \int \frac{\mathrm{d}p_{y}}{2\pi}
	\frac{p}{2\epsilon_0}
	r^{\rm TM}(\omega)
	e^{-2p z_a}
	\mathbf{\Pi}_+\mathbf{\Pi}_-
\end{align}
with $\epsilon_0$ the vacuum permittivity and $r^{\rm TM}$ the transverse magnetic reflection coefficient. The vectors $\mathbf{\Pi}_{\pm}=\mathbf{z}\mp \mathrm{i}\mathbf{p}/p$ describe the near-field polarization, where $\mathbf{p}$ is the component of the wave vector parallel to the surface ($q=p_{x}$ and $p=|\mathbf{p}|=\sqrt{p_x^2+p_y^2}$), and $\mathbf{z}$ the unit vector orthogonal to the surface.
For simplicity, we describe the metal using the spatially local Drude dielectric function
\begin{equation}
\epsilon(\omega)=1-\frac{\omega_{\rm p}^2}{\omega^2+\mathrm{i}\Gamma\omega},
\end{equation}
where $\omega_{\rm p}$ is the plasma frequency and $\Gamma$ the metal's dissipation rate.
In the near-field limit, the reflection coefficient can then be written as
\begin{equation}
r^{\rm TM}(\omega)
	=
	\frac{\epsilon(\omega)-1}{\epsilon(\omega)+1}
	=\frac{\omega_{\rm sp}^2}{
 	\omega_{\rm sp}^2-\omega^2-\mi\Gamma\omega},
\end{equation}
where we also defined the surface plasmon-polariton frequency $ \omega_{\rm sp}= \omega_{\rm p}/\sqrt{2}$ \cite{pitarke07}.
At low frequencies, we have $\mathrm{Im}\{r^{\rm TM}(\omega)\}\sim2\epsilon_0\rho\omega$, where $\rho=\Gamma/(\epsilon_0\omega_{\rm p}^{2})$ is the metal's resistivity.
Upon using the near-field expression of the Green tensor in Eq.~\eqref{lowvPLTE}, the power evaluates to
\begin{align}
P^{\mathrm{LTE}}
	\sim
	45\hbar
	\frac{v^4}{(2\pi)^3}
	\frac{\alpha_0^2\rho^2}{(2z_a)^{10}}>0,
\end{align}
i.e. a positive total power flux that tends to constantly increase the internal energy of the atom \cite{dedkov17}.
The previous value would have been even larger for a velocity-independent polarizability.

\subsection{Power and frictional force}

The balance between $P_{\rm in}$ and $P_{\rm out}$ is equivalent to the condition that in the NESS $\langle \dot{\mathbfh{d}}\cdot \mathbfh{E}\rangle=0$. 
Physically, this is equivalent to saying that the total power transferred to or dissipated within the atom must vanishes in the steady-state.
Proceeding as in Ref.~\cite{intravaia16} one can show that the condition $\langle \dot{\mathbfh{d}}\cdot \mathbfh{E}\rangle=0$ implies
\begin{multline}
\lim_{\substack{-t_{i}\to \infty\\ t\to \infty}}\mathrm{Re}\bigg(\frac{2\imath}{\pi}\int_{0}^{\infty}\mathrm{d}\omega\, \int_{0}^{t-t_{i}}\mathrm{d}\tau ~e^{-\imath \omega \tau}\int\frac{\mathrm{d}q}{2\pi}  \\
\times\mathrm{Tr}\left[\partial_{t}\underline{C}(t,t-\tau)\cdot \underline{G}^{\sf T}_{\Im}(q,\mathbf{R}_{a}, \omega)\right]
\\
e^{\imath q[x_a(t)-x_a(t-\tau)]}\bigg)=0,
\label{friction1}
\end{multline}
where $\tau=t-t'$ and $x_a(t)$ is the atomic trajectory. In the previous expression, $\underline{C}(t,t')=\langle \mathbfh{d}(t)\mathbfh{d}(t')\rangle$ is the dipole correlation matrix defined as in Refs.~\cite{intravaia16,intravaia16a}. 
Notice that, contrary to what was used in the main text, in this approach the usual (non-symmetric) quantum average is considered.

Using that $\partial_{t}=\partial_{\tau}$, we always have that in the NESS,  
\begin{multline}
\partial_{t}\underline{C}(t,t-\tau)\xrightarrow{\rm NESS} 
\\
\partial_{\tau}
\underline{C}(\tau)=\int \mathrm{d}\omega'\, (-\imath \omega')\underline{S}(\omega',v) e^{-\imath \omega'\tau},
\end{multline}
In the limit $-t_{i},t\to \infty$, Eq. \eqref{friction1} leads to the expression
\begin{equation}
2 \int_{0}^{\infty}\mathrm{d}\omega\,\int\frac{\mathrm{d}q}{2\pi} \,\omega_{q}^{-} \,
\mathrm{Tr}\left[\underline{S}^{\sf T}(-\omega_{q}^{-} ,v) \underline{G}_{\Im}(q,\mathbf{R}_{a}, \omega)\right]=0,
\label{friction2}
\end{equation}
where we used that the trace of the product of two Hermitian matrices is real.
The previous relation can be rewritten as follows: $P_{\rm rad}=-vF_{\rm fric}$, where
\begin{equation}
F_{\rm fric}=-2 \int_{0}^{\infty}\mathrm{d}\omega\,\int\frac{\mathrm{d}q}{2\pi}\, q
\mathrm{Tr}\left[\underline{S}^{\sf T}(-\omega_{q}^{-} ,v) \underline{G}_{\Im}(q,\mathbf{R}_{a}, \omega)\right]
\end{equation}
is the frictional force ($-vF_{\rm fric}$ is the work per unit of time preformed by the external agent) and
\begin{equation}
P_{\rm rad}=2 \int_{0}^{\infty}\mathrm{d}\omega\,\int\frac{\mathrm{d}q}{2\pi} \omega
\mathrm{Tr}\left[\underline{S}^{\sf T}(-\omega_{q}^{-} ,v) \underline{G}_{\Im}(q,\mathbf{R}_{a}, \omega)\right]
\end{equation}
defines the electromagnetic power dissipated (radiated) into the environment \cite{intravaia15a}. 

Importantly, in all these results the expression for the power spectrum tensor $\underline{S}$ are left unspecified and therefore they do not rely on any specific model for the atom's internal dynamics. 
For the case considered in the main text, $\underline{S}$ has the same expression as $\underline{\Sigma}$ where, however, the sign-function appearing in Eq. (4) is replaced by $2 \theta(\omega_{q}^{+})$, with $\theta(x)$ the Heaviside function \cite{intravaia19a}.

\section{Atomic Steady-state energy}
The equivalence of $P_{\rm in}$ and $P_{\rm out}$ leaves the energy of the atomic subsystem constant. It can be written as
\begin{equation}
\label{defE}
E
	=
	\lim_{\substack{t'\to t \\ t\to\infty}}
	\frac{\langle\dot{\mathbfh{d}}(t)\cdot\dot{\mathbfh{d}}(t')\rangle
	+\omega_a^2
	\langle\mathbfh{d}(t)\cdot\mathbfh{d}(t')\rangle}{2\alpha_0\omega_a^2}
	,
\end{equation}
where we highlighted the connection to the correlation function.
At late times, we already found that
\begin{align}
\mathbfh{d}(t)
	&=
	\int\frac{\mathrm{d}\omega}{2\pi}
	e^{-\mi\omega t}
	\int\frac{\mathrm{d}q}{2\pi}
	\underline{\alpha}(\omega,v)
	\mathbfh{E}_{0}(q,\mathbf{R}_{a},\omega_{q}^{+})
\end{align}
and similarly for its time derivative. 
Using the expression derived in the main text and the ones above, we have
\begin{align}
\label{intEnergy}
 E
 &=
 \mathrm{Tr}
 \int\frac{\mathrm{d}\omega}{2\pi}\hbar
\frac{\omega_a^2+\omega^2}{2\omega_a^2}
	\frac{\underline{\alpha}(\omega,v)
	\underline{\nu}(\omega,v)
	\underline{\alpha}^{\dag}(\omega,v)}{\alpha_0}
	\nonumber\\
	 &= 
 \int_{0}^{\infty}\mathrm{d}\omega\frac{\hbar}{2\pi}
\frac{\omega_a^2+\omega^2}{\omega_a^2}
	\mathrm{Tr}\left[\frac{\underline{\Sigma}(\omega,v)}{\alpha_0}\right],	
 \end{align} 
obtaining the definition for $\mathcal{E}(\omega,v)$ given in Eq.~(9). In the last expression we used the properties of the involved matrices discussed above, which also allow us to say that $\mathcal{E}$ (and therefore $E$) is an even function of $v$.

In equilibrium ($v=0$), we recover the FDT and then
\begin{align}
\label{Eequi}
 \mathcal{E}(\omega,0)=\frac{\hbar}{2\pi}
\frac{\omega_a^2+\omega^2}{\omega_a^2}
\mathrm{Tr}\left[\frac{\mathrm{Im}[\underline{\alpha}(\omega)]}{\alpha_0}\right],
 \end{align}
where $\underline{\alpha}_{\Im}(\omega,0)=\mathrm{Im}[\underline{\alpha}(\omega)]$. $E$ then takes the form previously obtained in the literature
\cite{Note1}.
For a generic atomic system, since $\underline{\alpha}_{I}(0)=0$, we have that $ \mathcal{E}(0,0)=0$ as reported in the main text.
Notice that in equilibrium, the BM approximation leads to a dipole correlation function given in terms of a (multi-)exponentially decay function \cite{lax63,mandel95,intravaia16}. 
For a single resonance $\omega_{a}$,
$\underline{C}^{\rm BM}(\tau)\approx\langle\mathbfh{d}\mathbfh{d}\rangle e^{-\mi \omega_{a}\tau-\gamma_{a} |\tau|}$,
where $\gamma_{a}\ge 0$ is related to the dissipative atom's dynamics. 
From Eq. \eqref{defE}, we have 
\begin{equation}
\mathcal{E}^{\rm BM}(\omega,0)\to \frac{\langle\mathbfh{d}^{2}\rangle}{2\pi \alpha_0}
\frac{\omega_a^2+\omega^2}{ \omega_a^2}
	\frac{\gamma_{a}}{(\omega-\omega_{a})^{2}+\gamma_{a}^{2}}
\end{equation}
(a multi-exponential decay leads to a similar expression). 
Therefore $\mathcal{E}^{\rm BM}(0,0)\not=0$ and, since $\langle\mathbfh{d}^{2}\rangle\propto \alpha_{0}$, its value only depends on the dissipative mechanism and the resonance. 
Usually, however, $\gamma_{a}=O(\alpha_{0}/[\epsilon_0\lambda^3])$ (e.g. radiative damping), with $\lambda$ a typical length scale characterizing the system's behavior (see also the main text). 
In this case, this property is inherited by $\mathcal{E}^{\rm BM}(0,0)$ which also becomes $O(\alpha_{0}/[\epsilon_0\lambda^3])$.

For $v\not=0$ neither $\underline{\alpha}$ nor $\underline{\nu}$ in Eq.~\eqref{intEnergy} are vanishing for $\omega=0$. Therefore $\underline{\Sigma}(0,v)\propto \alpha^{2}_{0}$ with a prefactor which is even in the velocity, leading to Eq.~(10) of the main text.
In general, the functional behavior of $E$ on $v$ depends on how the velocity relates to the other system's characteristic scales (e.g. $\omega_{a}$ and/or $\omega_{\rm sp}$), featuring a non-resonant and a resonant regime \cite{intravaia16}. Similar to the quantum frictional force, the energy changes from a power law to an exponential behavior (vanishing for decreasing $v$) as a function of the velocity.

\end{document}